\documentclass[traditabstract]{aa} 
\bibpunct{(}{)}{;}{a}{}{,}

\newcommand{\ergps}{erg\thinspace s$^{-1}$}

\newcommand{\ergpspsqcm}{erg\thinspace s$^{-1}$\thinspace cm$^{-2}$}

\newcommand{\psqcm}{cm$^{-2}$}
\newcommand{\nH}{$N_{\rm H}$}
\newcommand{\Msun}{$M_{\odot}$}

\usepackage[varg]{txfonts}
\usepackage{graphicx}
\usepackage{subfig}
\usepackage{color}

\begin{document}

\title{Steep-spectrum AGN in eROSITA Final Equatorial-Depth Survey (eFEDS): Their host galaxies and multi-wavelength properties}


\author{K. Iwasawa\inst{1,2}
  \and
  T. Liu\inst{3,4,5}
  \and
  Th. Boller\inst{3}
  \and
  J. Buchner\inst{3}
  \and
  J. Li\inst{6}
  \and
  T. Kawaguchi\inst{7}
  \and
  T. Nagao\inst{8}
  \and
  Y. Terashima\inst{8}
  \and
  Y. Toba\inst{8,9,10}
  \and
  J.~D.~Silverman\inst{11}
  \and
  R.~Arcodia\inst{3,12}
  \and
  Th. Dauser\inst{13}
  \and
  M. Krumpe\inst{14}
  \and
  K. Nandra\inst{3}
  \and
  J. Wilms\inst{15}
}

\institute{Institut de Ci\`encies del Cosmos (ICCUB), Universitat de Barcelona (IEEC-UB), Mart\'i i Franqu\`es, 1, 08028 Barcelona, Spain
         \and
         ICREA, Pg. Llu\'is Companys 23, 08010 Barcelona, Spain
         \and
         Max-Planck-Institut f\"ur extraterrestrische Physik, Giessenbachstra\ss e 1, 85748 Garching bei M\"unchen, Germany
         \and
         Department of Astronomy, University of Science and Technology of China, Hefei 230026, China
         \and
         School of Astronomy and Space Science, University of Science and Technology of China, Hefei 230026, China
         \and
         Department of Astronomy, University of Illinois at Urbana-Champaign, Urbana, IL 61801, USA
         \and
         Department of Economics, Management and Information Science, Onomichi City University, Onomichi, Hiroshima 722-8506, Japan
         \and
         Research Center for Space and Cosmic Evolution, Ehime University, Matsuyama, Ehime 790-8577, Japan
         \and
         National Astronomical Observatory of Japan, Mitaka, Tokyo 181-8588, Japan
         \and
         Institute of Astronomy and Astrophysics, Academia Sinica, Taipei 10617, Taiwan
         \and
         Kavli Institute for the Physics and Mathematics of the Universe, WPI, The University of Tokyo, Kashiwa, Chiba 277-8583, Japan
         \and
         MIT Kavli Institute for Astrophysics and Space Research, 70 Vassar Street, Cambridge, MA 02139, USA
         \and
         Universit\"at Erlangen/N\"urnberg, Dr.-Remeis-Sternwarte, Sternwartstra\ss e 7, 96049 Bamberg, Germany
         \and
         Leibniz-Institut f\"ur Astrophysik Potsdam (AIP), An der Sternwarte 16, 14482 Potsdam, Germany
         \and
         Dr. Karl Remeis-Sternwarte \& Erlangen Centre for Astroparticle Physics, Sternwartstra\ss e 7, 96049 Bamberg, Germany
         }


 
\abstract{We selected sources with a steep soft-X-ray-band spectrum with a photon index of $\Gamma >2.5$ ---measured by eROSITA on board the Spectrum-R\"ontgen-Gamma (SRG)--- from the eFEDS AGN catalogue as candidates of highly accreting supermassive black holes, and investigated their multi-wavelength properties. Among 601 bright AGN with 0.2-5 keV counts of greater than 100, 83 sources ($\approx 14$\%) are classified as steep-spectrum sources. These sources have typical 0.5-2 keV luminosities of $L_{\rm SX}\approx 10^{44}$ \ergps\ and the majority of them are found at redshifts below $z=1$. In comparison with sources with flatter spectra, these sources have, on average, a UV (or optical) to 2 keV luminosity ratio that is  larger by $\sim 0.3$ dex and bluer optical-to-UV continuum emission. They also appear to be radio quiet based on the detection rate in the FIRST and VLASS surveys. Their host galaxies ---at least in the redshift range of $z=0.2$-0.8, where the AGN--galaxy decomposition results from the Subaru Hyper Suprime-Cam imaging are available--- tend to be late-type and have smaller stellar masses ($M_{\star}\sim 10^{10.5}\, M_{\sun}$) than those of sources with flatter spectra. These properties are similar to those found in nearby narrow-line Seyfert 1 galaxies, in agreement with the picture that they are AGN with elevated accretion rates and are in the early growth phase of black hole and galaxy co-evolution.
However, the steep-spectrum sources are not exclusively narrow-line Seyfert 1 galaxies; indeed many are broad-line Seyfert 1 galaxies, as found by a catalogue search. This suggests that these steep-spectrum sources may be black holes generally with high accretion rates but of a wide mass range, including a few objects emitting at $L_{\rm SX}\geq 10^{45}$ \ergps, of which black hole masses can be close to $10^9$ \Msun.}

\titlerunning{Steep-spectrum AGN in eFEDS}
\authorrunning{K. Iwasawa et al.}
\maketitle
%

\section{Introduction}

The ROSAT All-Sky Survey \citep{voges1999} discovered many soft X-ray-bright active galactic nuclei (AGN), and these tend to be classified as narrow-line Seyfert 1 (NLS1) galaxies \citep{boller1996,laor1997}. Following black hole mass measurements of these AGN, a consensus formed that they are likely to harbour highly accreting, smaller black holes compared to broad-line Seyfert galaxies \citep[e.g.][]{grupe2004, xu2012}.
NLS1s have typical black hole masses of log $M_{\rm BH}/M_{\sun}\approx 6$-8 and Eddington ratios of $\lambda_{\rm Edd}\approx 0.1$-1 \citep[e.g.][]{jarvela2015,cracco2016,rakshit2017,paliya2023}. This means that a steep X-ray spectrum could be an indicator that an AGN has a high accretion rate \citep{brightman2013,risaliti2009,shemmer2008}. When an accretion rate approaches the Eddington ratio, an enhanced soft photon field from the accretion disc cools the hot corona, and optically thick Comptonised disc emission at the soft X-ray energies is elevated \citep[e.g.][]{done2012}. These would lead to a steep X-ray spectrum.

Decades later, eROSITA ---on board the Spectrum-R\"ontgen-Gamma \citep[SRG;][]{sunyaev2021,predehl2021,merloni2012}--- started to scan the whole sky in the X-ray band. With the large effective area in the soft X-ray band below 2 keV, eROSITA is expected to be sensitive to such steep-spectrum AGN and should therefore be a good tool with which to look for  highly accreting super-massive black holes. This means that, as well as objects similar to local NLS1s, more luminous counterparts at higher redshifts can also be detected in the eROSITA survey. An illustrative example has already been discovered: one of the Subaru high-$z$ exploration of low-luminosity quasars (SHELLQs), HSC\,J0921+0007 at $z=6.56$ \citep{matsuoka2018}, was detected by eROSITA and the subsequent Chandra observation revealed its steep spectrum of $\Gamma\simeq 3.2$ \citep{wolf2023}.

The eROSITA Final Equatorial-Depth Survey (eFEDS) is an early mission survey field of 142 deg$^2$ observed by eROSITA for science verification \citep{brunner2022}. AGN selected in eFEDS have been carefully matched with optical counterparts and multi-wavelength data \citep{salvato2022}. X-ray spectral analysis results on these AGN are presented by \citet{liu2022}. Much of the field has been covered by the Subaru Strategic Program survey \citep{aihara2018} with the Hyper Suprime-Cam \citep[HSC,][]{miyazaki2018}, which provides high-sensitivity optical imaging. Results of a decomposition of AGN and host galaxy for individual eROSITA-detected AGN at $z=0.2$-0.8, using the HSC imaging data, are available \citep{li2023}.

We selected steep-spectrum X-ray sources from bright sources in the eFEDS AGN catalogue in an attempt to separate out AGN with high accretion rates. We examined their host galaxies and multi-wavelength properties. We adopted the $\lambda $CDM cosmology with $H_0=70$ km\,s$^{-1}$\,Mpc$^{-1}$, and $\Omega_{\Lambda} = 0.7$.

\section{Selection of steep-spectrum AGN}

\subsection{Bright AGN sample}

We used the eFEDS AGN catalogue version 17.3 \citep{salvato2022}. Filtering with {\tt inArea90}=True, {\tt CTP\_quality} $>2$, and {\tt CTP\_redshift\_grade}$>2$  ensures the selection of sources that: are in regions with exposure of $>500$\,s ---these comprise 90\% of the area of eFEDS---, have a reliable optical counterpart; and a reliable redshift, respectively \citep[see][]{salvato2022,liu2022}. This filtering process is recommended by the eFEDS team, and yields 18659 sources. As we are going to select sources based on their spectral slopes, sources of interest need to have sufficient counts to enable a reasonable slope measurement. This requirement reduces the number of sources significantly. We adopted a selection threshold of greater than one hundred 0.2-5 keV counts; this leaves 601 sources, or 3.2\% of the detected AGN. This gives a spectral slope measurement in the 0.4-2.2 keV band with a typical uncertainty of $\simeq\pm 0.2$, as shown below. We note that, following inspection of the optical image, ID\,442 was suspected to be a star  and was therefore dropped.

We mainly used {\tt CTP\_redshift} ---the redshifts of the optical counterparts adopted in the catalogue \citep{salvato2022}--- for source redshifts. Spectroscopic measurements exist for 380 of the 601 bright AGN. The recent release of the SDSS DR18 provides further spectroscopic redshifts, which increases their total number to 520. Most of the new spectroscopic redshifts agree with the photometric estimates of {\tt CTP\_redshift} within $\pm 0.2$ but about 4\% (25) deviate by 0.87-4.1, mostly to lower values; their redshifts and luminosities have been revised, although they represent a minority and have little impact on the population statistics, which is the main focus of this work. 

\begin{figure} \centerline{\includegraphics[width=0.4\textwidth,angle=0]{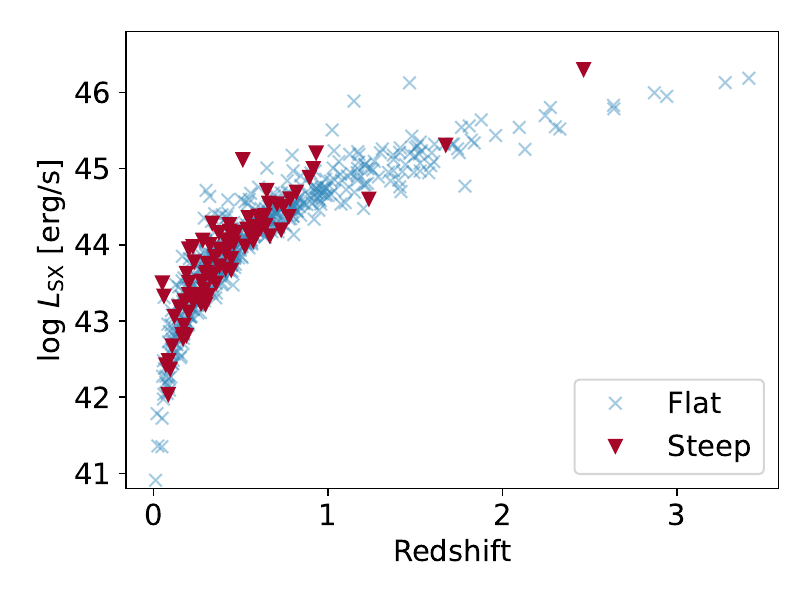}}
\caption{Plot of soft X-ray luminosity against redshift for the flat- and steep-spectrum sources (in blue crosses and red triangles, respectively). }
\label{fig:zlx}
\end{figure}

These bright AGN are distributed over redshifts up to $z\sim 3.5$ and have soft X-ray (SX: 0.5-2 keV) luminosities ($L_{\rm SX}$) of up to log $L_{\rm SX}$ = 46.3 [\ergps], as shown in Fig. \ref{fig:zlx}. The median value of the soft X-ray luminosities is log $\tilde L_{\rm SX} = 44.03$ (Table 1).

\begin{figure} 
\centerline{\includegraphics[width=0.4\textwidth,angle=0]{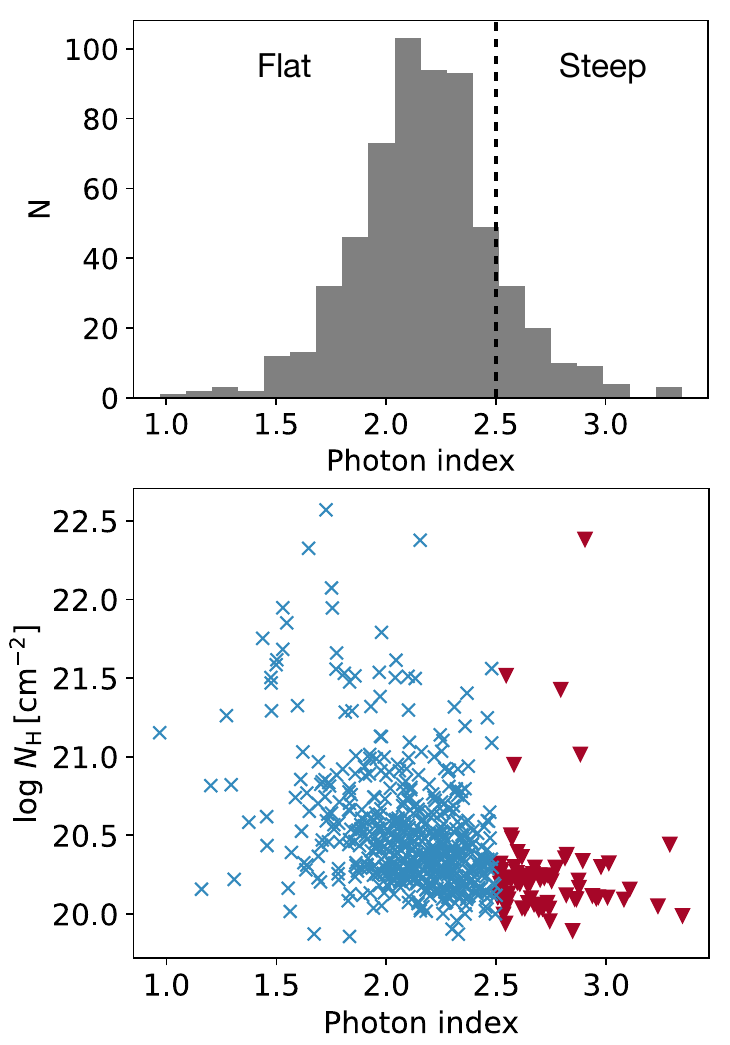}}
\caption{Distribution of the soft X-ray photon index and absorbing column of the whole bright AGN sample of 601 sources. Upper panel: Photon index distribution. The vertical dashed line indicates the boundary of $\Gamma = 2.5$ between the flat- and steep-spectrum sources. Lower panel: Flat- and steep-spectrum sources (symbols are the same as in Fig. 1) in the photon-index--log\,$N_{\rm H}$ plane.}
\label{fig:gamma_dist}
\end{figure}

Steep-spectrum sources were selected based on the spectral fits presented by \citet{liu2022}. We used the results obtained for the 0.4-2.2 keV spectra with an absorbed power-law fit. This band is the most sensitive band of eROSITA and represents the soft-X-ray portion of the spectrum well; it is not affected by any spectral hardening that could occur at higher energies. This band brackets the conventional 0.5-2 keV (SX) band with a small margin added on either end to ensure good flux measurements in the SX band \citep{liu2022}. The distribution of the photon index ($\Gamma $) ---represented by the posterior median \citep[{\tt Gamma\_s\_med},][]{liu2022} of the fit--- for the 601 bright sources is shown in Fig. \ref{fig:gamma_dist}.

The photon indices are distributed from 1.0 to 3.3 and have both median and mean values of 2.18. We classified 83 sources with $\Gamma > 2.5$ as steep-spectrum sources; these amount to 14\% of the bright AGN sample. The choice of the slope threshold is arbitrary but the small proportion of 14\% indicates that the selected steep-spectrum sources occupy a peripheral part of the slope distribution (Fig. \ref{fig:gamma_dist}). The typical 68\% error of the slope for the selected sources is $\pm 0.2$ and 96\% (81) of the steep-spectrum  sources have errors of smaller than $0.27$, justifying the adopted source-counts threshold for delivering a reasonable accuracy in spectral-slope measurements. The majority of the steep-spectrum sources (78) have a logarithmic absorbing column density log \nH\ of smaller than 20.5 [\psqcm ] (see Fig. \ref{fig:gamma_dist}), indicating that there is little effect of degeneracy between $\Gamma $ and \nH\ for obtaining the steep slopes in the sample.

Sources with $\Gamma \leq 2.5$ are, for simplicity, referred to as flat-spectrum sources hereafter. The flat-spectrum sources include a relatively increased number of obscured sources ---with absorbing column densities of up to log\,\nH $\simeq 10^{22.5}$ [\psqcm] (Fig. \ref{fig:gamma_dist})--- when compared to the steep-spectrum sources: the proportion of sources with log\,\nH $> 20.5$ [\psqcm] is about 40\%.

\begin{table}
\caption{Bright AGN sample from eFEDS}
\label{tab:sample}
\centering
\begin{tabular}{cccccc}
  \hline\hline
Class & $\Gamma $ & $N$ & $\tilde z$ & log$\tilde F_{\rm SX}$ & log$\tilde L_{\rm SX}$\\
  \hline
  Steep & $>2.5$ & 83 & 0.376 & $-13.00$ & 43.95 \\
  Flat & $\leq 2.5$ & 518 & 0.465 & $-12.98$ & 44.06 \\[5pt]
  Total & --- & 601 & 0.452 & $-12.98$ & 44.03\\
\hline
\end{tabular}
\tablefoot{Class: X-ray spectral class; $\Gamma $ photon index measured in the 0.2-2 keV band; $N$: number of sources; $\tilde z$: median redshift; $\tilde F_{\rm SX}$: median absorption-corrected soft-X-ray flux in units of \ergpspsqcm; $\tilde L_{\rm SX}$: median soft X-ray luminosity in units of \ergps.}
\end{table}

\subsection{Steep-spectrum source fraction}

\begin{figure}
  \centerline{\includegraphics[width=0.4\textwidth,angle=0]{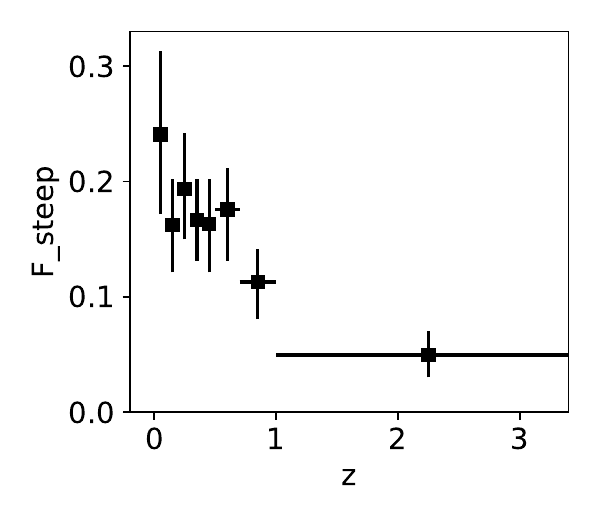}}
\caption{Steep-spectrum source fraction ($F_{\rm steep}$) as a function of redshift. The error bars of each $F_{\rm steep}$ indicate the 68\% compatible interval. }
\label{fig:fsteep_z}
\end{figure}

The fraction of the steep-spectrum sources in the bright AGN sample is 14\%. The fraction at the lowest redshift range ($z<0.2$) is $\sim 20$\%, which decreases as a function of increasing redshift down to $\sim 5$\% at $z>1$ (Fig. \ref{fig:fsteep_z}). This decrease is partly due to the bandpass effect. X-ray spectra of AGN below 2-3 keV are generally steeper than that in the harder band. As the redshift increases, the steep part of the rest-frame spectrum goes out and the flatter spectrum in the rest-frame hard band enters the band.

In the low count regime, obtained spectral parameters tend to be dictated by the prior: in the case of the spectral slope, the prior is $\Gamma = 2$ \citep{liu2022}, which is in the range of the flat-spectrum sources. Our selection of bright sources with counts greater than 100 means that the decreasing quality of the spectra for sources at higher redshift does not seem to be the cause of the declining steep-spectrum source fraction, as the median source counts at the three redshift ranges ---168 cts at $z<0.5$, 152 cts at $z=0.5$-1, and 134 cts at $z=1$-3.5--- vary by only $\approx 20$\%.

\section{Host galaxy properties}

\begin{figure}
  \centerline{\includegraphics[width=0.5\textwidth,angle=0]{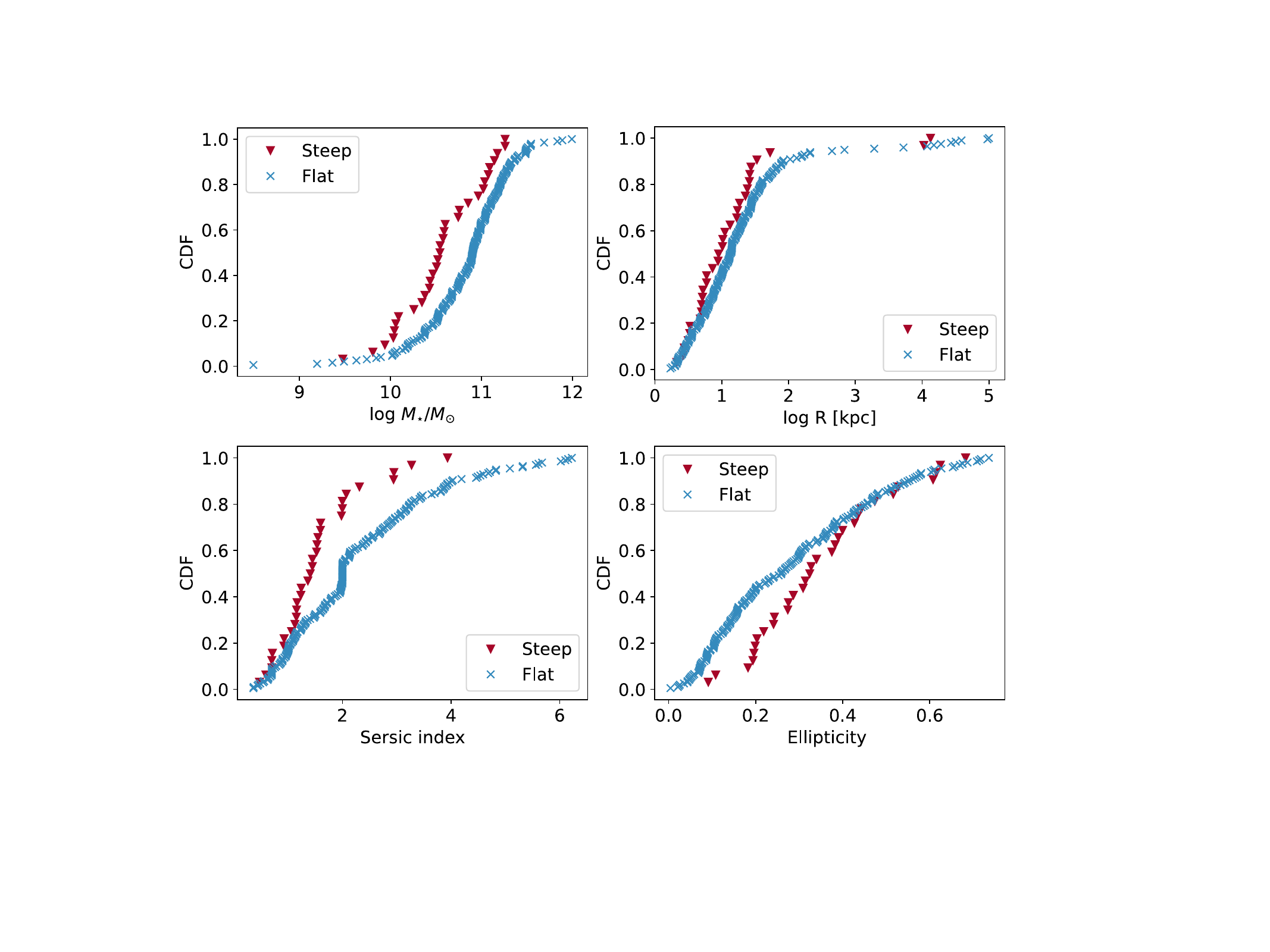}}
\caption{Cumulative density functions of host galaxy properties computed separately for steep- and flat-spectrum sources. Upper-left: Stellar mass $M_{\star}$ in units of solar masses. Upper-right: Galaxy size log $R$ in kpc. Lower-left: S\'ersic index. Lower-right: Ellipticity. Symbols are the same as in Fig. 1.}
\label{fig:host_CDFs}
\end{figure}

The two-dimensional imaging decomposition of host galaxy light from AGN was performed for 3796 AGNs in eFEDS at redshifts $z = 0.2$-0.8 \citep{li2023}, where the image quality of the Subaru HSC data permits it. This limited redshift range results in our bright AGN sample size being reduced to 229. The number of steep-spectrum sources is 32, which is 14\% of the total sample and is also the same proportion as that in the total bright AGN sample.

We compare stellar mass, log $M_{\star}$, galaxy size (half-light radius of the semi-major axis), log\,$R$ [kpc], the S\'ersic index, $n$ and the ellipticity, $\epsilon$, obtained from the decomposition and the S\'ersic profile fits between the two samples.
The cumulative density functions (CDFs) of these features for the two samples are shown in Fig. \ref{fig:host_CDFs}. The distribution of log\,$R$ has a long tail towards larger values, which likely result from poor decomposition, but the other parameters are not too far from the normal distribution with variable degrees of tails. The Student-{\it t} distribution has an extra parameter to a normal distribution to accommodate tails and is relatively robust against outliers. We compared the posterior mean value of each feature of the host galaxies between the flat- and steep-spectrum sources using the BEST: Bayesian Estimation Supersedes the {\it t}-test \citep{kruschke2013} method, implemented with PyMC3.

\begin{figure}
\centerline{\includegraphics[width=0.5\textwidth,angle=0]{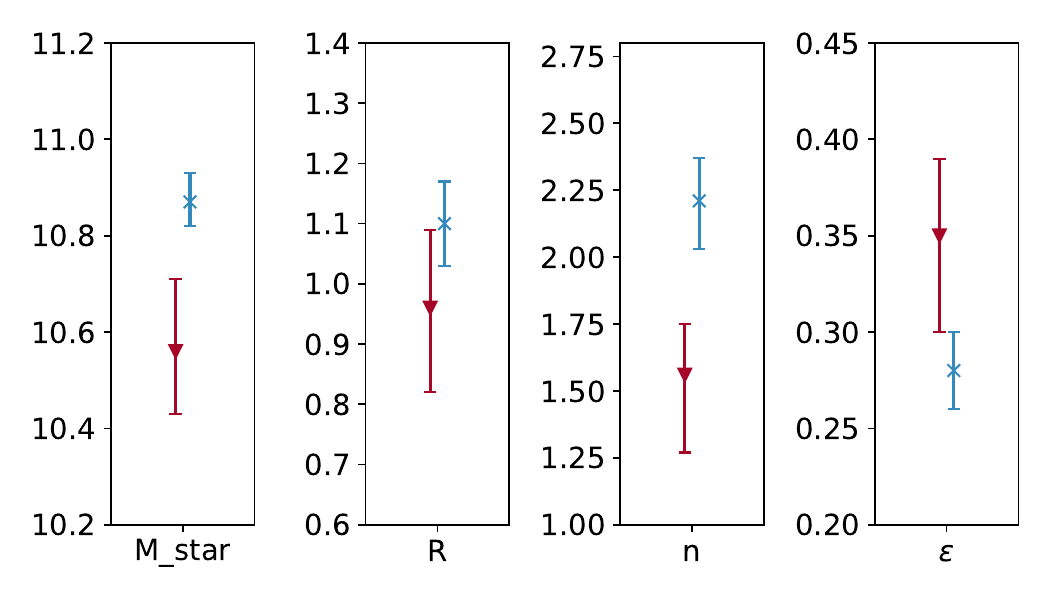}}
\caption{The mean values of the steep- and flat-spectrum sources for the four host galaxy parameters, log $M_{\star}$ in units of $M_{\sun}$, the galaxy size, log $R$, in kpc, S\'ersic index, $n$, and the ellipticity, $\epsilon $. Symbols are the same as in Fig. 1. The error bars show the 89\% highest-density interval of the respective mean values.}
\label{fig:hostparams}
\end{figure}

The results are shown in Fig. \ref{fig:hostparams}. We also computed the posteriors of the difference between the means of the two samples. The probabilities that the differences are non-zero are 99.9\% for stellar mass, 86\% for galaxy size, $\approx 100$\% for S\'ersic index, and 97\% for ellipticity. The above results show that the steep-spectrum sources are found in host galaxies with significantly smaller $M_{\star}$ and $n$ than flat-spectrum sources.

As mentioned in \citet{li2023}, $n$ in some galaxies was fixed at 0.7 or 2.0 in the decomposition fitting, when it did not converge to reasonable values. There are 4 (2 and 2 for $n=0.7$ and $n=2.0$, respectively) cases in steep-spectrum sources and 22 (5 and 17 for $n=0.7$ and $n=2.0$, respectively) in flat-spectrum sources (see Fig. \ref{fig:host_CDFs}). When these sources are excluded, the significant difference in $n$ between the flat-spectrum and steep-spectrum samples still has a mean value of 1.6 (the 89\% compatible interval: 1.3-1.8) for steep-spectrum sources and 2.3 (2.1-2.5) for flat-spectrum sources, in agreement with the original result (see Fig. \ref{fig:hostparams}).

We carefully examined whether or not the above differences are artefacts of systematic errors in the optical image decomposition, as the steep-spectrum sources are mostly unobscured (Sect. 2.1) and are therefore expected to be bright in optical with respect to host galaxy light. An over-subtraction of the AGN component could cause both smaller $M_{\star}$ and a flatter galaxy profile. The differences do not appear to be an effect of nuclear obscuration within the flat-spectrum sample alone, as no statistical differences in $M_{\star}$ and $n$ were found between obscured sources  (log\,\nH\,$\geq 20.5$ [\psqcm]) and unobscured sources  (log\,\nH\,$<20.5$ [\psqcm ], which match the steep-spectrum
sources). The differences between the steep- and flat-spectrum samples in terms of $M_{\star}$ and $n$ persist when the samples are filtered by various matched properties:  optical magnitudes in $r$ and $g$, X-ray flux, and X-ray luminosity. We also investigated the accuracy of redshifts, which could also affect host galaxy property measurements. All the redshifts are flagged as {\tt CTP\_redshift\_quality} of either 4 ---that is, two photometric estimate methods, LePHARE \citep{ilbert2006} and DNNz (Nishizawa et al. in prep.) agree--- or 5, where a spectroscopic redshift is available, indicating reasonable accuracy. The number of steep-spectrum sources with spectroscopic redshifts is 21 out of 32. The results do not change even when using only sources with spectroscopic redshifts or only low-redshift sources with $z<0.5,$ for which better accuracy in an AGN--galaxy decomposition is expected. Furthermore, these steep-spectrum sources follow the average trend seen in the optical-to-X-ray-luminosity and AGN-fraction-to-X-ray-luminosity diagrams of \citet[][see their Fig. 5]{li2023}, indicating no evidence for over-subtraction of the AGN component in the decomposition. Therefore, the differences in the host galaxy properties are unlikely to be caused by the systematic error. Given the small sample size and the reliability of the measurements of parameters such as $n$, we refrain from drawing any strong conclusions as to the differences in the host galaxy properties at present, but our results nevertheless warrant further discussion (Sect. 6).



\section{Multi-wavelength properties}

\subsection{Optical/UV to X-ray luminosity ratio}

\begin{figure}
  \centerline{\includegraphics[width=0.5\textwidth,angle=0]{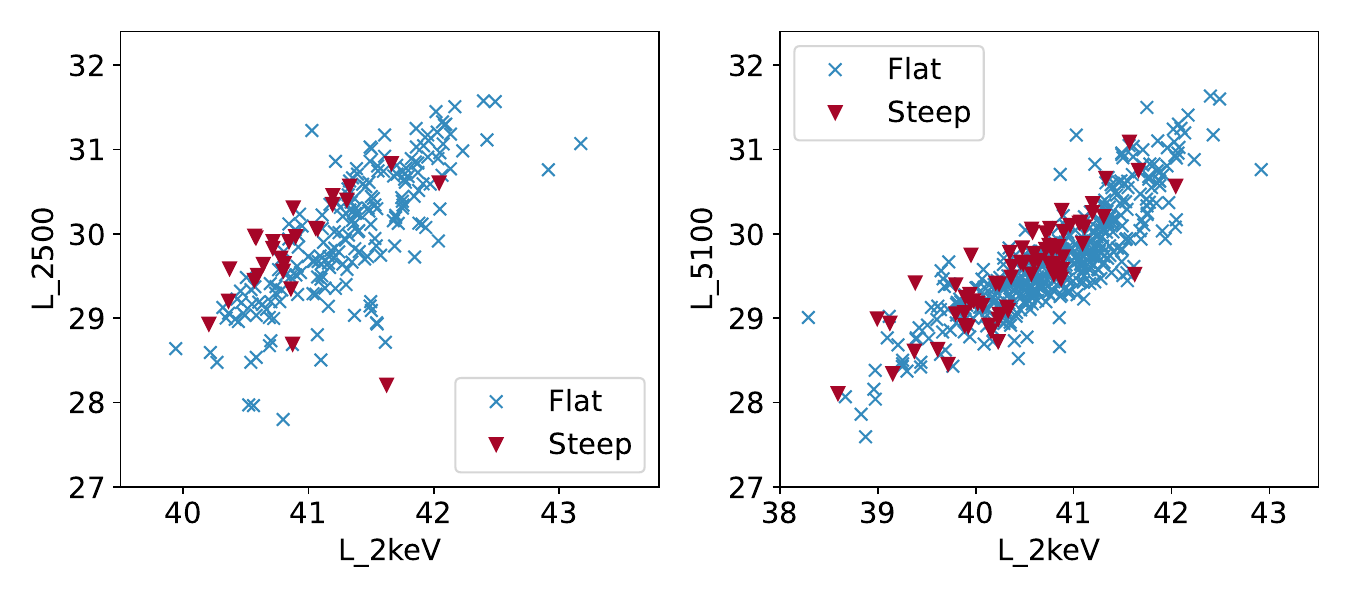}}
\caption{Optical/UV to X-ray luminosity relation between the flat- and steep-spectrum sources. The rest-frame 2500\r{A} against 2 keV luminosity (left) and the rest-frame 5100\r{A} against 2 keV luminosity (right) are plotted. Symbols are the same as in Fig. 1.}
\label{fig:Luv-L2}
\end{figure}

We used the rest-frame monochromatic luminosities at 2500\r{A} and 5100\r{A} and the narrow-band (2 eV width) 2 keV luminosities contained in the eFEDS AGN catalogues \citep{salvato2022,liu2022} to make the correlation diagrams of $L_{2500}$--$L_2$ and $L_{5100}$--$L_2$  (Fig. \ref{fig:Luv-L2}).
The steep-spectrum sources are found to lie at the upper envelope of both correlations, indicating that the steep-spectrum sources show a larger $L_{2500}$ or $L_{5100}$ at a given $L_2$ compared to the flat-spectrum sources.
The $t$-test using the BEST method shows that the mean log\,$(L_{2500}/L_2)$ of the steep-spectrum sample is larger than that of the flat-spectrum sample by 0.31 (the 89\% compatible interval of 0.19-0.42). For log\,$(L_{5100}/L_2)$, the mean difference is found to be 0.21 (0.15-0.28). 


\subsection{Radio detection rate}

We compared radio detection rates between the steep- and flat-spectrum samples, using the radio sources catalogued in the VLA FIRST survey at 1.4 GHz \citep{becker1995} and the Very Large Array Sky Survey (VLASS) at 3 GHz \citep{gordon2021}. The eFEDS is not yet covered by the LOFAR Two-Metre Sky Survey (LoTSS) as of Data Release 2 \citep[DR2,][]{hardcastle2023}. We used the 14dec2017 version of the FIRST survey catalogue via its search facility\footnote{https://sundog.stsci.edu/cgi-bin/searchfirst} and the VLASS Quick Look and Single Epoch Catalog (Epoch 1) on VizieR. The optical counterpart positions of the eFEDS sources from the DESI Legacy Imaging Survey DR8 \citep[LS8,][]{dey2019} were used to match the radio sources with a search radius of 2\arcsec. There are five and three matches out of the 83 steep-spectrum sources with the FIRST and VLASS sources, respectively, while there are 83 (FIRST) and 77 (VLASS) matches out of the 518 flat-spectrum sources. Matched radio sources have total fluxes of 0.7-625 mJy at 1.4 GHz and 1-1600 mJy at 3 GHz. The detection rates are $6\pm 3$\% and $4\pm 3$\% for steep-spectrum sources and $16\pm 2$\% and $15\pm 2$\% for flat-spectrum sources in the FIRST and VLASS surveys, respectively. The steep-spectrum sources appear to be less frequently detected in both radio frequencies than the flat-spectrum sources, although the sample size is small and therefore the result remains inconclusive.

All three of the steep-spectrum sources detected in VLASS are found among the five detected in FIRST. The two brightest radio sources (eFEDS ID\,71 and 834) with 17 mJy and 75 mJy at 1.4 GHz, respectively, have been classified as BL Lacs, and were also detected previously in X-ray with ROSAT \citep{plotkin2008,massaro2009,veron-cetty2006}. Others detected at a few mJy are a broad-line Seyfert 1 galaxy \citep[BLS1; ID\,231,][]{oh2015} and two galaxies (ID\,139 and 477).

\subsection{WISE colours}

\begin{figure}
    \centerline{\includegraphics[width=0.35\textwidth,angle=0]{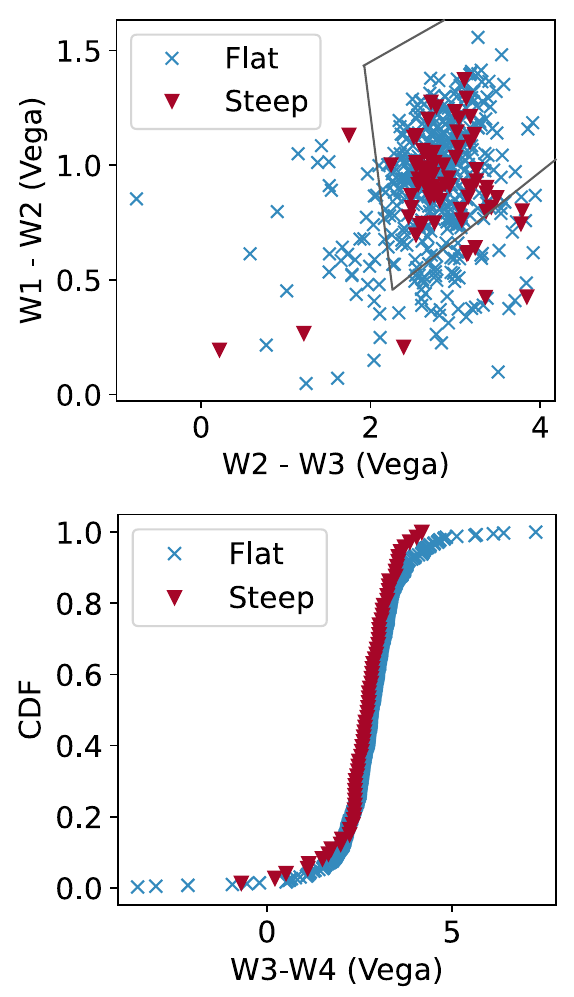}}
  \caption{Infrared colours of the steep- and flat-spectrum sources. Symbols are the same as in Fig. 1. Upper: WISE colour--colour diagram of W1$-$W2 against W2$-$W3. The AGN wedge proposed by \citet{mateos2012} is drawn as a grey solid line. Lower: Cumulative density function of W3$-$W4 of flat-spectrum and steep-spectrum sources.}
\label{fig:WISEcolcol}
\end{figure}

We used sources detected in all four WISE bands ---namely 424 flat-spectrum and 74 steep-spectrum sources--- to investigate the WISE colours. To compare with the earlier study of AGN WISE colours, we converted the WISE photometry in the eFEDS catalogue to Vega magnitude. The (W1$-$W2)--(W2$-$W3) colour--colour diagram is shown in Fig. \ref{fig:WISEcolcol}. While the W1$-$W2 colour of flat-spectrum sources spreads wider than that of steep-spectrum sources, the colour distributions of both samples align to each other and are compatible with the general AGN colours: W1$-$W2 $> 0.8$ \citep{stern2012,yan2013} or the AGN wedge \cite{mateos2012} shown in Fig. \ref{fig:WISEcolcol}. There is no systematic difference in the W3$-$W4 colour between the two samples either (Fig. \ref{fig:WISEcolcol}).

\subsection{Rest-frame mid-IR to UV SEDs}

\begin{figure*}
  \centerline{\includegraphics[width=0.85\textwidth,angle=0]{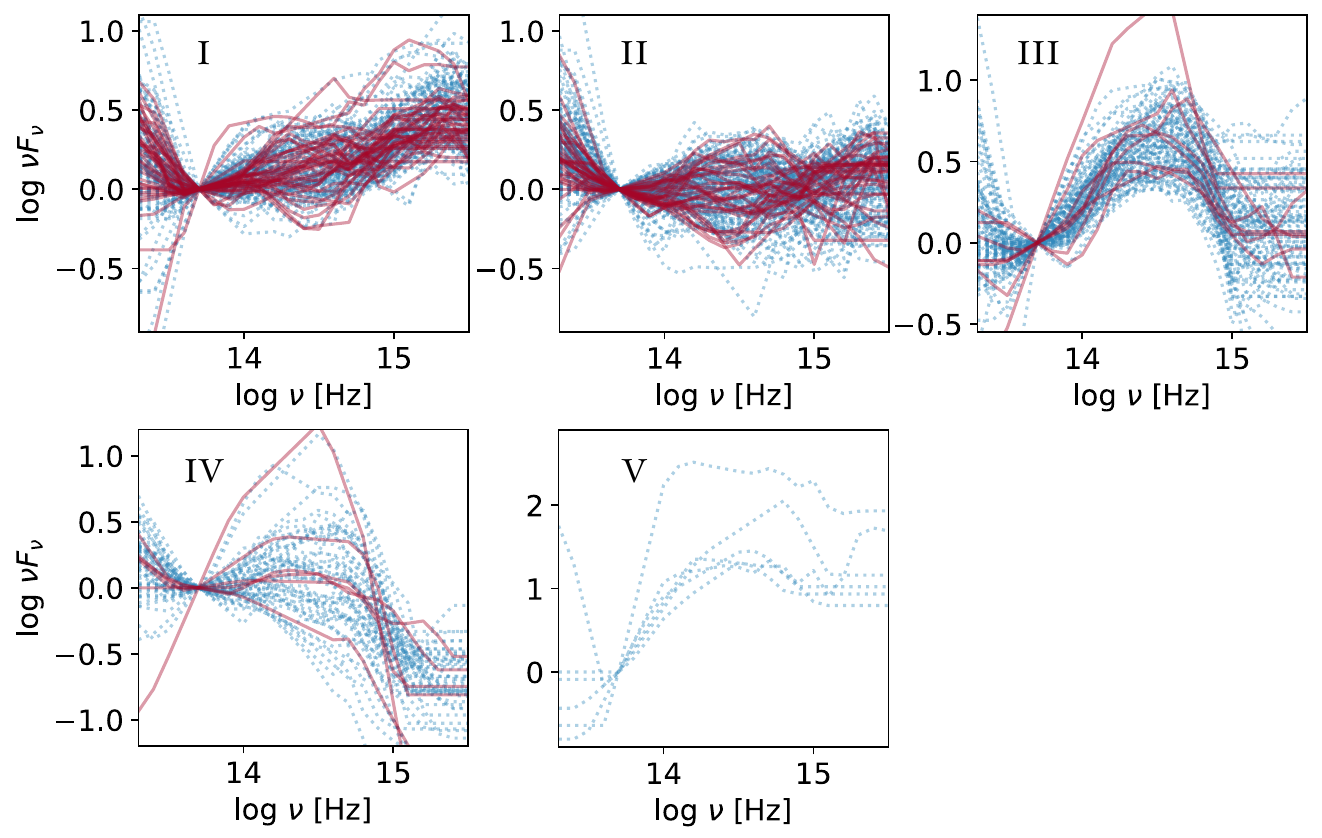}}
\caption{Five types of normalised SEDs, according to the clustering analysis (Table \ref{tab:SEDclusters}). The flat- and steep-spectrum sources are plotted in blue dotted and red solid lines, respectively.}
\label{fig:SEDclusters}
\end{figure*}

\begin{figure}
  \centerline{\includegraphics[width=0.48\textwidth,angle=0]{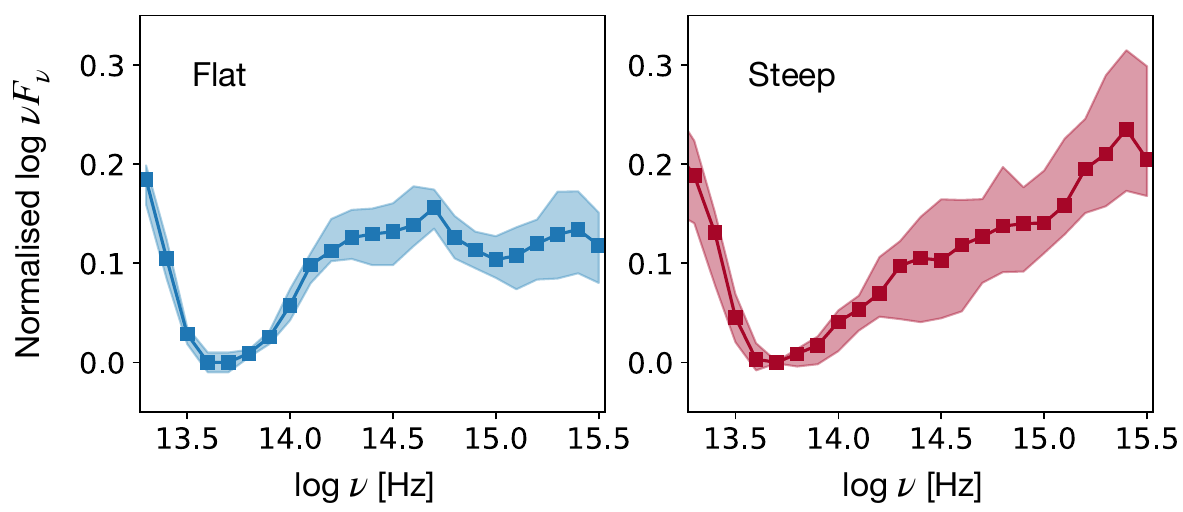}}
\caption{Stacked rest-frame SEDs of the individual sources in Fig. \ref{fig:SEDclusters}. The shaded area indicates the 68\% compatible intervals computed by bootstrapping. Left: Flat-spectrum sources at $z<1$. Right: Steep-spectrum sources.}
\label{fig:stackSEDs}
\end{figure}

We used multi-wavelength photometry (in up to 15 bands) from mid-IR (WISE W4) to UV (GALEX FUV) available in the catalogue \citep{salvato2022} to construct a rest-frame spectral energy distribution (SED) for each source in the bright AGN sample. All the photometric data have been corrected for the Galactic extinction \citep{salvato2022}. We adopted a median value when multiple photometric measurements are available in a given band from different telescopes. As most of the steep-spectrum sources are at $z<1$, we used 426 flat-spectrum sources at $z<1$  to cover the same rest-frame wave bands in order to make a comparison.

For each source, available photometric data were converted to Janskys and then into units of $\nu f_{\nu }$. Twenty-three equally spaced logarithmic intervals between $10^{13.3}$ Hz  and $10^{15.5}$ Hz were set for the rest-frame frequency. After SED data points at the 23 frequencies were estimated by linear interpolation in logarithmic space, they were normalised at 6\,$\mu $m (log $\nu = 13.7$ [Hz]). These SEDs come in various shapes but they can be largely divided into four categories with a few outliers, as shown by the following clustering analysis.


We applied the basic $k$-means method \citep{lloyd1982,forgy1965} implemented in {\tt scikit-learn} \citep{scikit-learn} to all of the 509 SEDs and the number of clusters was chosen by analysing the silhouette coefficient \citep{rousseeuw1987} for a variable number of clusters, which was found to be five. The five clusters of SEDs (named I, II, III, IV, and V) are shown separately in Fig. \ref{fig:SEDclusters}. The following characteristics are observed in each class of SEDs. Cluster I shows a blue excess towards UV, or a `big blue bump', characteristic of the disc emission of unobscured AGN. Cluster II shows  a flat spectrum throughout from mid-IR to UV. Cluster III has a relative excess peaking around $1\,\mu $m (log $\nu\sim 14.5$ [Hz]), suggesting dominant stellar emission from a host galaxy. Cluster IV shows a suppression of optical to UV light, indicating obscuration to a nuclear source, and Cluster V shows six outliers with large optical to $6\mu $m ratios.
The numbers of flat- and steep-spectrum sources in each SED class are given in Table \ref{tab:SEDclusters}. 
The above characteristics of Clusters I and II fit those of unobscured AGN and the majority of the steep-spectrum sources (69 out of 83) are indeed found in those clusters. The SEDs of Clusters III and IV are likely attributable to obscured AGN, and these clusters are mainly populated by flat-spectrum sources, as expected. Among the steep-spectrum sources, the two BL Lacs (Sect. 4.2) and two sources with \nH\ in excess of $10^{21}$ \psqcm\ are found in Cluster III. The origin of the SEDs for the remaining ten steep-spectrum sources found in Clusters III and IV is unclear, despite the fact that they all have reliable optical counterparts ({\tt CTP\_quality} of 4) and reliable redshifts ({\tt CTP\_redshift\_grade} of 4 or 5, apart from one). These sources are typically low-luminosity objects with a median log $L_{\rm SX}=43.3$ [\ergps ] at low redshift ($\tilde z\simeq 0.26$). Their soft spectrum cannot be attributed to galaxy emission because the luminosities are too high and must originate from AGN. The steep, soft X-ray spectra could result from a dusty warm absorber, which would redden the optical-UV light.

The average rest-frame SEDs of the flat- and steep-spectrum sources constructed by median stacking are shown in Fig. \ref{fig:stackSEDs}. The blue end of the SEDs of some obscured objects lack data, namely above $10^{15}$~Hz. This part of the SED for these sources was estimated by extrapolating a power-law fitted to the three bluest available data points before computing the median at each frequency. The 68\% compatible intervals obtained from bootstrapping are shown by the shaded area. The SED shape in the mid-IR part is similar in the two spectrum classes, while the near-IR to UV part differs, which reflects their distribution into the different SED clusters discussed above. The average SED of the steep-spectrum sources is dominated by the monotonically rising, unobscured AGN emission towards the UV band, while that of the flat-spectrum sources, in contrast, shows a flatter spectrum in the near-IR to UV range, which is a result of the mixture of various types, including obscured objects.

\begin{table}
\caption{Clustering of mid-IR to UV SEDs}
\label{tab:SEDclusters}
\centering
\begin{tabular}{ccccc}
  \hline\hline
Cluster & Flat & Steep & Total & $f_{\rm S}$ \\
  \hline
  I & 156 & 33 & 189 & 0.18 (0.15-0.20) \\
  II &  121 & 36 & 157 & 0.23 (0.20-0.26)\\
  III & 92 & 8 & 100 & 0.08 (0.06-0.12) \\
  IV & 51 & 6 & 57 & 0.11 (0.08-0.16)\\
  V & 6 & 0 & 6 & 0.09 (0.02-0.23) \\
\hline
\end{tabular}
\tablefoot{Cluster: SED cluster (see Fig. \ref{fig:SEDclusters}); Flat: number of flat-spectrum sources; Steep: number of steep-spectrum sources; Total: sum of the steep- and flat-spectrum sources; $f_{\rm S}$: proportion of the steep-spectrum sources represented by the posterior median. The 68\% compatible interval is shown in parenthesis.}
\end{table}

\section{High-luminosity steep-spectrum sources}

There are five steep-spectrum sources with $L_{\rm SX}$ exceeding $10^{45}$ \ergps: ID 11, 40, 351, 421, and 644. The bolometric correction for this X-ray luminosity range is $\kappa\equiv L_{\rm bol}/L_{\rm X}\sim 60$, where $L_{\rm bol}$ and $L_{\rm X}$ are bolometric and 2-10 keV luminosities, respectively \citep{marconi2004}. When $L_{\rm X} = \eta L_{\rm SX}$, their black hole mass ($M_{\rm BH}$) can be expressed as $M_{\rm BH}\approx 5\times 10^8\,M_{\sun}\,\eta\lambda_{\rm Edd}^{-1}(\kappa /60)(L_{\rm SX}/10^{45}\,{\rm erg\,s}^{-1})$. The luminosity conversion factor, $\eta$, is 1.2 or 0.55 when the X-ray spectrum is described by a single power-law of $\Gamma = 2.0$ or 2.5, respectively. However, as the spectrum usually flattens above 2 keV, $\eta $ probably remains close to unity for many sources. If the Eddington ratio, $\lambda_{\rm Edd}$, is significantly lower than 1, the $M_{\rm BH}$ of these high-luminosity sources can be as massive as $\sim 10^{9}M_{\sun}$. However, only two of them have spectroscopic redshifts of $z<1$. The  source with the highest redshift, ID 421, has $z= 2.466$ from LePHARE, which is adopted here but does not agree with $z=0.47$ from DNNz. A spectroscopic redshift measurement is required to verify the high-redshift and high-luminosity nature of this source, as well as the other high-luminosity candidates with photometric redshift estimates.

\section{Discussion and summary}

We selected 83 AGN with a steep-spectrum ($\Gamma > 2.5$) in the soft-X-ray band from the bright eFEDS AGN sample of 601 objects with detected counts of $>100$ in an attempt to identify AGN accreting at a high Eddington rate. In a comparison with flat-spectrum sources, the steep-spectrum sources show some statistical differences in terms of the  properties of their host galaxy, their optical-UV-to-X-ray ratio, and their  radio detection rate.

When a specific accretion rate is defined by log\,$(L_{\rm SX}/M_{\star})$ as a relative measure of $\lambda_{\rm Edd}$, the steep-spectrum sources have a larger mean specific accretion rate than the flat-spectrum sources by 0.2 (the 89\% compatible interval of 0.05-0.37) dex, as found by the BEST evaluation. This supports the idea that the steep-spectrum sources have, on average, elevated accretion rates relative to flat-spectrum sources. The larger UV-optical-to-X-ray ratio in the steep-spectrum sources compared to the flat-spectrum sources (Sect. 4.1) agrees with the known trend of steepening optical-to-X-ray spectral slope $\alpha_{\rm OX}$ with increasing $\lambda_{\rm Edd}$ found in a previous study \citep[e.g.][]{lusso2010}. The results of our rest-frame SED study verify that the steep-spectrum sources generally show blue optical-to-UV continuum, which has been found in ROSAT-found NLS1s with steep spectra \citep{grupe2004}. The radio detection rate of the steep-spectrum sources may be lower than that of the flat-spectrum sources. Although further tests with a larger sample are needed for verification, the possible radio quietness of the steep-spectrum sources is consistent with the general trend of an inverse correlation between radio-loudness and the Eddington ratio in AGN \citep{ho2002,yang2020}.

The steep-spectrum sources, on average, reside in late-type galaxies with smaller $M_{\star}$ (Sect. 3), but these tendencies need to be verified with a larger sample. This is reminiscent of NLS1s, which have been found to reside mainly in spiral-type host galaxies, possibly with enhanced bars \citep{krongold2001,crenshaw2003,ohta2007,orbandexivry2011,mathur2012,jarvela2018}, including radio-loud NLS1s \citep{olguin-iglesias2020,varglund2022}.


All these properties of the steep-spectrum sources are compatible with those of NLS1s, which are considered to be highly accreting (or rapidly growing) black holes hosted by galaxies in the early phase of their build-up \citep[e.g.][]{grupe2004,mathur2000}. Whether or not a steep X-ray spectrum is synonymous with an NLS1 classification is an open question. We explored this possibility by matching the X-ray sources with catalogued NLS1s. We used the NLS1 catalogue of \citet{paliya2023}, which was produced by analysing optical spectra from the SDSS DR17. These latter authors also released a catalogue of broad-line Seyfert 1 galaxies (BLS1s) from the same analysis. The redshift range of the catalogued objects is limited below $z=0.8$ or $z=0.9$, depending on the spectrograph used to ensure the rest-frame wavelength coverage of their interest, but the majority of the steep-spectrum sources lie in this redshift range. It should be noted that galaxies that exhibit optical spectra typical of Seyfert 2 or LINER-type optical emission lines or of normal galaxies with no apparent AGN signatures are not contained in these catalogues.


\begin{table}
\caption{Matches with NLS1s and BLS1s.}
\label{tab:x-opt_classes}
\centering
\begin{tabular}{lccc}
  \hline\hline
& NLS1s & BLS1s & Total \\
  \hline
  Flat & 25 & 145 & 170 \\
  Steep & 14 & 23 & 37 \\[5pt]
  Total & 39 & 168 & 207 \\
\hline
\end{tabular}
\tablefoot{NLS1s and BLS1s classified by \citep{paliya2023} using the SDSS DR17 optical spectra were matched with the flat- and steep-spectrum sources.}
\end{table}

Both flat- and steep-spectrum sources were matched with the NLS1 and BLS1 catalogues using the optical counterpart (LS8) positions with a search radius of 2\arcsec. Matched numbers are given in Table \ref{tab:x-opt_classes}. As the choice of $\Gamma >2.5$ for steep-spectrum sources is arbitrary, it is not surprising to see some NLS1s among the flat-spectrum sources. However, steep-spectrum sources are not always NLS1s. The narrowness of the permitted lines that define a NLS1 is likely linked to a small black hole mass. On the other hand, if the steep X-ray spectral slope primarily results from a large $\lambda_{\rm Edd}$, the nature of the X-ray spectrum is independent of black hole mass and steep-spectrum sources can be NLS1s or BLS1s, as long as their $\lambda_{\rm Edd}$ is large. Some of the high-luminosity steep-spectrum sources discussed in Sect. 5 could correspond to large black holes still operating at high accretion rates. Despite the presence of a number of BLS1s among the steep-spectrum sources, a significant proportion ($\sim 40$\% in Table \ref{tab:x-opt_classes}) of NLS1s ---which are likley to harbour small black holes--- might lead to the host galaxy properties of the steep-spectrum sources being similar to those of NLS1s.


The theory governing the growth of supermassive black holes  predicts that black holes with super-Eddington accretion rates will be more abundant at higher redshifts and suggests that this evolution may be observable \citep{kawaguchi2004,shirakata2019}. This is contrary to the apparent decrease in the proportion of steep-spectrum sources beyond $z\sim 1$  (Sect. 2.2). If this decrease towards higher redshifts were due to the redshift effect, as discussed in Sect. 2.2, the steep part of the X-ray spectrum of AGN with high accretion rates would need to be limited below the rest-frame 2 keV. However, the spectra of bright NLS1s are generally steeper than those of normal Seyferts at energies even above 2 keV, which is the case for the eROSITA-detected $z\sim 6.5$ quasar HSC~J0921+0007 \citep{wolf2023}, for example, which has $\Gamma\sim 3$ in the rest-frame energies above $\sim 5$ keV. This discrepancy needs to be further investigated using a larger sample such as that being gathered by the eROSITA All-Sky Survey \citep{merloni2012,predehl2021}.

\begin{acknowledgements}
This work is based on data from eROSITA, the soft X-ray instrument aboard SRG, a joint Russian-German science mission supported by the Russian Space Agency (Roskosmos), in the interests of the Russian Academy of Sciences represented by its Space Research Institute (IKI), and the Deutsches Zentrum f\"ur Luft- und Raumfahrt (DLR). The SRG spacecraft was built by Lavochkin Association (NPOL) and its subcontractors, and is operated by NPOL with support from the Max Planck Institute for Extraterrestrial Physics (MPE). The development and construction of the eROSITA X-ray instrument was led by MPE, with contributions from the Dr. Karl Remeis Observatory Bamberg and ECAP (FAU Erlangen-Nuernberg), the University of Hamburg Observatory, the Leibniz Institute for Astrophysics Potsdam (AIP), and the Institute for Astronomy and Astrophysics of the University of T\"ubingen, with the support of DLR and the Max Planck Society. The Argelander Institute for Astronomy of the University of Bonn and the Ludwig Maximilians Universit\"at Munich also participated in the science preparation for eROSITA. Funding for the Sloan Digital Sky Survey IV has been provided by the Alfred P. Sloan Foundation, the U.S. Department of Energy Office of Science, and the Participating Institutions. Funding for the Sloan Digital Sky Survey V has been provided by the Alfred P. Sloan Foundation, the Heising-Simons Foundation, the National Science Foundation, and the Participating Institutions. SDSS acknowledges support and resources from the Center for High-Performance Computing at the University of Utah. The SDSS web site is \url{www.sdss.org}.
SDSS is managed by the Astrophysical Research Consortium for the Participating Institutions of the SDSS Collaboration, including the Carnegie Institution for Science, Chilean National Time Allocation Committee (CNTAC) ratified researchers, the Gotham Participation Group, Harvard University, Heidelberg University, The Johns Hopkins University, L’Ecole polytechnique f\'{e}d\'{e}rale de Lausanne (EPFL), Leibniz-Institut f\"{u}r Astrophysik Potsdam (AIP), Max-Planck-Institut f\"{u}r Astronomie (MPIA Heidelberg), Max-Planck-Institut f\"{u}r Extraterrestrische Physik (MPE), Nanjing University, National Astronomical Observatories of China (NAOC), New Mexico State University, The Ohio State University, Pennsylvania State University, Smithsonian Astrophysical Observatory, Space Telescope Science Institute (STScI), the Stellar Astrophysics Participation Group, Universidad Nacional Aut\'{o}noma de M\'{e}xico, University of Arizona, University of Colorado Boulder, University of Illinois at Urbana-Champaign, University of Toronto, University of Utah, University of Virginia, Yale University, and Yunnan University. The Hyper Suprime-Cam (HSC) collaboration includes the astronomical communities of Japan and Taiwan, and Princeton University. The HSC instrumentation and software were developed by the National Astronomical Observatory of Japan (NAOJ), the Kavli Institute for the Physics and Mathematics of the Universe (Kavli IPMU), the University of Tokyo, the High Energy Accelerator Research Organization (KEK), the Academia Sinica Institute for Astronomy and Astrophysics in Taiwan (ASIAA), and Princeton University.
This research made use of NASA/IPAC Extragalactic Databases (NED), which is funded by the NASA and operated by the California Institute of Technology, the VizieR catalogue access tool, and the SIMBAD database, both operated at CDS, Strasburg, France. Software packages of Astropy, Astroquery, HEASoft~6.27.1, Matplotlib, PyMC, Numpy, Scipy, Pandas and IPython were used for data analysis. KI acknowledges support by the Spanish MCIN under grant PID2019-105510GB-C33/AEI/10.13039/501100011033 and ``Unit of excellence Mar\'ia de Maeztu 2020-2023'' awarded to ICCUB (CEX2019-000918-M). MK acknowledges support from DLR grant FKZ\,50\,OR\,2307.
\end{acknowledgements}

\bibliographystyle{aa} \bibliography{efeds}{}

\end{document}